\documentclass[runningheads]{llncs}
\usepackage{multirow}
\usepackage{graphicx}
\usepackage{booktabs} 
\usepackage{array}
\usepackage[colorlinks=true]{hyperref}
\usepackage[ruled,vlined,linesnumbered]{algorithm2e}
\usepackage{multicol}
\usepackage{cite}
\usepackage{lscape}

\makeatletter

\usepackage{bibentry}

\begin{document}
\setcounter{secnumdepth}{3}

\setlength{\tabcolsep}{5pt}

\title{
WikiCSSH: Extracting and Evaluating Computer Science Subject Headings from Wikipedia\thanks{This paper is an extended version of the peer-reviewed workshop paper: Han, K., Yang, P., Mishra, S., \& Diesner, J. (2020). WikiCSSH: Extracting Computer Science Subject Headings from Wikipedia \cite{WikiCSSH2020Han}. In Proceedings of Scientific Knowledge Graphs Workshop co-located with 24th International Conference on Theory and Practice of Digital Libraries. This material is based upon work supported by the Korea Institute of Science and Technology Information under Grant No. C17031. We thank Kehan Li for assistance with data annotation, and anonymous reviewers for their feedback.}
}

\titlerunning{WikiCSSH}
%
%
\author{Kanyao Han\inst{1} \and
Pingjing Yang\inst{1} \and
Shubhanshu Mishra\inst{1} 
\and
Jana Diesner\inst{1}}
\authorrunning{K. Han et al.}

\institute{School of Information Sciences,\\ University of Illinois at Urbana-Champaign, Champaign, IL, 61820, USA \email{kanyaoh2@illinois.edu;py2@illinois.edu;\\mishra@shubhanshu.com;jdiesner@illinois.edu}}
\maketitle              
\begin{abstract} 
Hierarchical domain-specific classification schemas (or subject heading vocabularies) are often used to identify, classify, and disambiguate concepts that occur in scholarly articles. In this work, we develop, apply, and evaluate a human-in-the-loop workflow that first extracts an initial category tree from crowd-sourced Wikipedia data, and then combines community detection, machine learning, and hand-crafted heuristics or rules to prune the initial tree. This work resulted in WikiCSSH; a large-scale, hierarchically-organized vocabulary for the domain of computer science (CS). Our evaluation suggests that WikiCSSH outperforms alternative CS vocabularies in terms of vocabulary size as well as the performance of lexicon-based key-phrase extraction from scholarly data. WikiCSSH can further distinguish between coarse-grained versus fine-grained CS concepts. The outlined workflow can serve as a template for building hierarchically-organized subject heading vocabularies for other domains that are covered in Wikipedia.

\end{abstract}

\section{Introduction} 
A scholarly publication can be considered as a collection of concepts or terms. Identifying these concepts allows us to build better search interfaces\footnote{\href{https://www.nlm.nih.gov/bsd/pubmed.html}{https://www.nlm.nih.gov/bsd/pubmed.html}}, study temporal trends in the evolution of concept usage \cite{Mishra2016a,Packalen2019,Nielsen2017Scholia}, compute conceptual expertise of authors \cite{Mishra2018Expertise}, and study citation patterns in scholarly data \cite{Mishra2018SelfCite,levine2010rankings}, among other practical applications. For many domains, e.g., biomedicine, mathematics, and physics, well curated, controlled, and structured vocabularies have been developed, which are commonly referred to as subject heading vocabularies (or simply, subject headings). These subject headings index relevant concepts in a domain, and organize these concepts into a hierarchical structure (e.g., concepts and sub-concepts), which facilitates coarse-grained and fine-grained knowledge organization that represents the breadth and depth of a field. Examples of prominent vocabularies are Medical Subject Headings (MeSH)\footnote{\href{https://www.nlm.nih.gov/mesh/concept\_structure.html}{https://www.nlm.nih.gov/mesh/concept\_structure.html}}, Physics subject headings (PhySH)\footnote{\href{https://physh.aps.org/}{https://physh.aps.org/}}, and Mathematics Subject Classification (MSC)\footnote{\href{https://mathscinet.ams.org/msc/msc2010.html}{https://mathscinet.ams.org/msc/msc2010.html}}.

In the domain of computer science (CS), a commonly used vocabulary is the ACM Computing Classification System (ACM CCS)\footnote{\href{https://dl.acm.org/ccs}{https://dl.acm.org/ccs}}. While this vocabulary has been curated by CS domain experts, it is being updated more slowly than the field advances, and is comparatively small-scale: The latest version of ACM CCS was released in 2012 (and its predecessor in 1998) and contains about 2,000 subject headings, while MeSH is updated once a year and contains 25,000 subject headings. Furthermore, the ACM CCS schema contains coarse-grained concepts that are helpful for identifying and categorizing relatively broad research areas of computing, but is not designed to also capture concrete, fine-grained concepts. To remedy these shortcomings, recently, the Computer Science Ontology (CSO) \cite{salatino2018computer} has been introduced as an automatically constructed ontology. CSO was extracted from scholarly papers, contains more than 14,000 subject headings and semantic relations between them, and can help to identify both broad and narrow research areas of computing. However, CSO fails to distinguish between core CS concepts versus CS-related concepts that emerge at the nexus of CS and other fields through interdisciplinary work. In this paper, we refer to these related concepts as ancillary CS concepts. Examples of ancillary CS concepts include ``gender" and ``aircraft". Another strength of CSO is that it links each concept to multiple knowledge bases, including Wikidata and Freebase. However, CSO does not yet leverage the vast amount of human effort used to organizing knowledge in Wikipedia. To address the outlined limitations, we herein report on the extraction and evaluation of a large-scale, hierarchical, and semi-curated CS vocabulary that distinguishes between coarse-grained and fine-grained concepts as well as between core and ancillary CS concepts. The resulting vocabulary is furthermore grounded in knowledge provided by many people over time in the form of the Wikipedia Category Tree (WCT)\footnote{\href{https://en.wikipedia.org/wiki/Special:CategoryTree}{https://en.wikipedia.org/wiki/Special:CategoryTree}}. We refer to our resulting vocabulary as \textit{Wikipedia-based Computer Science subject headings (WikiCSSH)}\footnote{\href{https://github.com/uiuc-ischool-scanr/WikiCSSH}{https://github.com/uiuc-ischool-scanr/WikiCSSH}} \cite{WikiCSSH}. WikiCSSH was created with a mixed methods approach to extracting CS-relevant subject headings, which included breadth first search in the WCT; followed by manual filtering, community detection, embedding-based classification, and human-created rules for removing false positives. The construction of WikiCSSH benefited from the automatic association of Wikipedia pages with Wikipedia categories, which we used for the automatic expansion of WikiCSSH to include pages affiliated with Wikipedia categories into WikiCSSH. Finally, our comparative analysis shows that none of currently existing vocabularies, which were built with different approaches (i.e., expert curation for ACM CCS, data mining of scholarly work for CSO, and crowd-sourcing for WikiCSSH) is complete or can comprehensively outperform any other vocabulary.

Our project makes four main contributions. First, we provide a large hierarchical subject headings schema for CS with more than 700,000 CS concepts that are divided into core and ancillary concepts. Second, our work shows how to leverage the Wikipedia Category Tree for this purpose. This methodology might serve as a template for the construction of vocabularies for other domains for which information is available from Wikipedia. Third, this paper illustrates the challenges resulting from using Wikipedia data for this specific task, shows solutions to these challenges, and implements a workflow with human-in-the-loop processes to overcome some of these hurdles. Finally, this paper evaluates and analyzes the quality of various vocabularies in terms of concept coverage patterns and key-phrase extraction performance, which further indicates the advantages and disadvantages of various vocabulary building approaches, including curation by human experts, data mining, and crowd-sourcing (Wikipedia) based approaches.

\section{Related Work} 
Various domains and scholarly communities have developed their own hierarchical, domain-specific vocabularies, such as MeSH for biomedicine. MeSH is particularly useful for practical applications due to its hierarchical and non-cyclical nature. Furthermore, MeSH, along with MEDLINE, an annotated biomedical corpus, can be used to track the evolution of biomedical concepts over time and to create concept profiles of authors \cite{Mishra2016a,Packalen2019,Nielsen2017Scholia}. The fields of mathematics and physics also have developed domain-specific vocabularies, namely, Mathematics Subject Classification (MSC) and Physics subject headings (PhySH). Besides these domain specific solutions exists the Wikipedia Category Tree (WCT), which covers a large number of domains and is used to classify Wikipedia articles. WCT is curated by the Wikipedia community. For CS, ACM CSS \cite{osborne2015klink} and CSO \cite{salatino2018computer} are the two prominent controlled vocabularies. A comparison of various domain-specific and cross-domain controlled vocabularies is shown in table \ref{tab:cs_category_stats}.

\begin{table}[!htbp]
    \centering
    \caption{Comparison of existing controlled vocabularies for various domains.}
    \begin{tabular}{lll
    >{\raggedright\arraybackslash}p{0.39\textwidth}
    >{\raggedright\arraybackslash}p{0.16\textwidth}
    }
    \toprule
    \textbf{Name}      & \textbf{Type} & \textbf{Size} & \textbf{Curation}                          & \textbf{Domain}  \\ \midrule
    \textbf{MeSH}      & Fine grained  & 25K           & National Library of Medicine               & Biomedicine      \\
    \textbf{PhySH}     & Fine grained  & 3.5K             & Americal Physical Society                  & Physics          \\
    \textbf{PACS}      & Subject level & 9.1K          & American Institute of Physics              & Physics          \\
    \textbf{MCS}       & Subject level & 6.1K          & Mathematical Reviews and Zentralblatt MATH & Mathematics      \\
    \textbf{CCS}   & Subject level & 2K          & Association of Computer Machinery          & Computer Science \\
    \textbf{WCT} & Fine grained  & 1M+           & Wikipedia contributors                     & Open domain      \\
    \bottomrule
    \end{tabular}
    \label{tab:cs_category_stats}
\end{table}

While expert-constructed vocabularies often trade off size for quality and accuracy, automatically generated vocabularies often flip this relationship. Constructing vocabularies from structured, crowd-sourced data has become a viable approach \cite{medelyan2008topic,shang2018automated,lehmann2015dbpedia,hoffart2013yago2,wang2010timely} to boost both size and quality. For example, prior research has leveraged Wikipedia as a comprehensive knowledge base \cite{medelyan2008topic,shang2018automated}, e.g., for building multilingual DBpedia \cite{lehmann2015dbpedia} and temporal YAGO2 \cite{hoffart2013yago2,wang2010timely}. Since Wikipedia and the referenced related projects are not domain-specific to CS, we herein aim to leverage Wikipedia to develop a domain-specific, hierarchical, and non-cyclical vocabulary that distinguishes between coarse-grained and fine-grained concepts as well as between core and ancillary CS concepts, and a methodology to achieve these goals.

\section{Methods}
\subsection{Wikipedia Category Tree}

As of April 2020, the Wikipedia Category Tree (WCT) consists of 1.6M categories with 10.9M inter-category links and 217.6M category-page links. Each category in the WCT can have multiple parents as well as multiple children. Links between categories are referred as parent-child links. Each category has multiple affiliated pages. We assume that pages affiliated with a category refer to concrete concepts within that category. In other words, a category is a coarse-grained term that refers to a relatively broad research area or topic, while a page is a fine-grained term that refers to a relatively narrow research area, topic, or concept within a category. It is important to note that WCT is not necessarily a tree as it contains circular paths, e.g., \emph{Mathematics} $\rightarrow$ \emph{Philosophy of mathematics} $\rightarrow$ \emph{Formalism} $\rightarrow$ \emph{Formal sciences} $\rightarrow$ \emph{Mathematics} $\rightarrow$ \emph{Philosophy of mathematics}. Furthermore, since WCT is crowd-sourced and open-domain, it contains many parent-child relationships which are not relevant for our task of identifying categories relevant to CS research concepts. For example, in the parent-child chain \emph{Computing and society} $\rightarrow$ \emph{Social media} $\rightarrow$ \emph{Fiction about social media}, the category \emph{Computing and society} is relevant to CS in our context, but the category \emph{Fiction about social media} is not. Furthermore, \emph{Fiction about social media} leads to additional irrelevant categories (such as \emph{Novels about social media}), and this pattern is recursive.

\subsection{Building an initial CS domain-specific subtree}

To construct a CS specific subject headings schema, we started by extracting an initial CS subtree (ICS) as described next (see Algorithm \ref{alg:workflow}). The following categories were chosen as starting points because they represent five highest-level domains relevant to CS: \emph{computer science}, \emph{information science}, \emph{computer engineering}, \emph{statistics}, and \emph{mathematics}. These five categories constitute the first level of our initial CS subtree, and determine the overall breadth of our vocabulary. We recursively updated ICS with all children of the categories in the current ICS using a breadth first search over WCT. Redundant categories were removed during this search since we removed all categories based on exact matches of phrases that have occurred before. This resulted in an ICS with more than 1.4 million categories, which were organized in 20 levels (depth of ICS). Overall, the extraction process performed in this first step has resulted in high recall but low precision for CS-relevant categories.

\begin{algorithm}[!htbp]
\DontPrintSemicolon
\SetAlgoLined
\SetKwInOut{Input}{input}
\SetKwInOut{Output}{output}
\SetKwData{WikiCSSH}{WikiCSSH}
\SetKwData{WCT}{WCT}
\SetKwData{ICS}{ICS}
\SetKwData{children}{categories}
\SetKwData{childrenNew}{$new_{cats}$}
\SetKwData{embeddingModels}{models}
\SetKwFunction{FindChildren}{Children}
\SetKwFunction{ManualFiltering}{ManualFiltering}
\SetKwFunction{CommunityDetection}{FindCommunities}
\SetKwFunction{FilterCommunities}{FilterCommunities}
\SetKwFunction{ExtractPages}{ExtractPages}
\SetKwFunction{TrainModels}{TrainModels}
\SetKwFunction{FilterICS}{Filter}
\Input{\WCT, $\ICS \leftarrow$ {Initial Categories}, $rules$}
\Output{\WikiCSSH}
\begin{multicols}{2}
$\childrenNew \leftarrow \ICS$\;
\While{$\childrenNew \neq \emptyset$}{
$\children \leftarrow \FindChildren{\childrenNew}$\;
$\childrenNew \leftarrow \children - \ICS$\;
$\ICS \leftarrow \ICS \cup \childrenNew$\;
}
$\ICS \leftarrow \FilterICS{\ICS, manual}$\;
$communities \leftarrow \CommunityDetection{\ICS}$\;
$\ICS \leftarrow \FilterICS{\ICS, communities}$\;
$models \leftarrow \TrainModels{\ICS}$\;
$\ICS \leftarrow \FilterICS{\ICS, models}$\;
$\ICS \leftarrow \FilterICS{\ICS, rules}$\;
$\WikiCSSH \leftarrow \ExtractPages{\ICS}$\;
\end{multicols}
 \caption{Building WikiCSSH}
 \label{alg:workflow}
\end{algorithm}

\subsection{Removing false positives from the ICS}
Our manual inspection of this ICS revealed a few major issues. 

First, as described above, we identified many categories that were not related to the domain of CS and should therefore be removed. These categories often appear in lower levels of our tree where the inclusion of even a single irrelevant category can lead to the inclusion of a large number of that category's irrelevant children. Second, while some categories were related to CS, a few of them were not useful for our intended use, i.e., building a structured vocabulary of subject headings relevant for tagging terms and indexing research in CS. These included names of CS conferences, researchers, and CS research/teaching institutes. We consider the two above-mentioned issues as cases of false positives, and describe our approach for removing those in Algorithm \ref{alg:workflow}. It is important to note that here, false positives and irrelevant categories are meant with respect to our purpose, not noise or irrelevance in Wikipedia itself. We fully acknowledge that any of the instances that we did not include in WikiCSSH might very well be excellent categories for other contexts and applications.

\subsubsection{Manual annotation for first three levels:}
The first three levels of the ICS contained a variety of broad, important sub-domains that are relevant to CS, such as \emph{artificial intelligence} and \emph{algorithms and data structures}. Considering that any false negatives and false positives in these levels can lead to a lack of relevant CS sub-domains or the inclusion of core research areas from other domains, respectively, we decided to manually annotate all categories in the first three levels (a total of 759) for relevance for our purpose. Based on that, we removed 259 (32\%) categories from the first three levels. Even though we also removed the children of these 259 categories, there were still around 1.4 million categories remaining in the ICS.

\subsubsection{Community detection:} 

A network with an inherent community structure can be grouped into sets (communities) of nodes such that each set is densely connected within, and weakly connected across communities~\cite{girvan2002community}. In our remaining ICS, categories from the same or similar domains or sub-domains were densely connected through child-parent links, such that we can assume CS-relevant categories to cluster. Considering the large size of the remaining ICS (1.4 million categories), we leveraged a widely used and fast community detection algorithm, namely, the Louvain algorithm \cite{blondel2008fast}. This algorithm identified a total of 288 clusters in the remaining ICS. The largest and smallest clusters contained 243,597 categories and 1 category, respectively, and the mean and median size of these 288 clusters were 5044 and 41, respectively. To identify and remove CS-irrelevant clusters, we utilized the overlap of categories in those clusters with terms in ACM CSS and CSO. Based on that, we removed 261 (94.1\%) clusters with a total of 0.4 million (28.6\%) categories that had no overlap with ACM CCS or CSO. Our inspection of the remaining 1 million categories showed that there were still substantial numbers of false positive categories. To address this issue, we next trained a machine learning model to predict false positive categories.

\subsubsection{Embedding-based classification:}

Our next step for reducing false positives was to use a machine learning model to automatically distinguish relevant from irrelevant categories with high accuracy. We utilized embedding based approaches, namely Elmo \cite{Peters2018}, poincare \cite{Nickel2017}, and node2vec \cite{Grover2016} embeddings, to capture the contextual information of the texts of Wikipedia categories, the structured information in our subtree, and the graph information of the child-parent links in our data. While we were able to create features through embeddings, it was difficult to obtain a training set with balanced labeled responses. Since the ratio of positive to negative categories in the remaining ICS was smaller than 1\%, we were not able to label enough positive instances for model training through annotating a sample from the remaining ICS. In view of this difficulty, we considered a total of 1756 categories that were shared between the remaining ICS and ACM CSS or CSO as positive responses. Next, we obtained negative responses by manually annotating a sample from the remaining ICS, and collecting the children of the annotated negative responses. Since we obtained tens of thousands of CS-irrelevant categories (negative responses), we randomly sampled about 1756 categories from them to create a balanced training set. We then utilized a multi-layer perceptron (MLP) to train a model to predict whether a category is CS-relevant or not. The cross validated (k = 10) F1 score of the model was around 90\%. The Elmo-based model performed best, and the addition of node2vec features improved the performance slightly (1\% to 2\%). Therefore, we utilized the MLP model based on the features from Elmo and node2vec to predict whether a category is relevant to CS or not. We then applied the trained model to the remaining ICS, and removed all categories labeled as CS-irrelevant from the remaining ICS. Also, if any category was classified as CS-irrelevant, we removed its child categories. This step removed the majority of categories from the ICS. The remaining ICS contained about 11,000 (1.1\%) categories out of all categories in our initial data.

\subsubsection{Human-created rules} 

After inspecting the remaining ICS, we still found a substantial number of false positive categories in it. We also observed that there were more false positive categories in the bottom levels than those in the top levels. Since manually identifying and removing individual categories is time-consuming, we developed a set of rules or heuristics to handle patterned cases of false positives that were not captured by any of the above-mentioned pruning steps. In order to find effective rules, we randomly sampled hundreds of categories from the remaining ICS, and manually annotated whether they were relevant or not. This in-depth work revealed that most false positive categories had common parent categories, and these parent categories often shared common patterns. For example, a commonly shared parent of false positive categories was \emph{Classification system by subject}. This category did not refer to classification methods or systems in CS, but classification schemas in other domains. We also found that the suffix \emph{by subject} in parent categories often led to the inclusion of false positive children categories into the remaining ICS as well. Another example of patterned false positives was \emph{$\langle$brand name$\rangle$ software}, which is relevant to CS in general, but irrelevant for our purpose. Therefore, we removed all categories containing the suffix \emph{by subject} and the prefix \emph{$\langle$brand name$\rangle$}. Using a process of filtering out false positive categories from the sample and locating their parents by tracing bottom-up parenthood links, we identified around 50 patterns, and created corresponding rules to remove them. Overall, we removed about 4000 (35\%) from the remaining ICS, and obtained 7355 categories. At this point, we had used 0.45\% of the categories from WCT for WikiCSSH.

\subsection{Extracting fine-grained terms} 

Since a CS subject headings schema should also contain fine-grained concepts within each research area, we utilized all of the pages affiliated with CS-relevant categories identified through the previous steps. Based on our assumption that pages inherit the characteristic of CS-relevance from categories with which they are affiliated, we recursively extracted pages which were relevant to CS. This step refined our WikiCSSH with 761,383 pages that were affiliated with the 7355 categories in our remaining ICS.

\subsection{Final WikiCSSH}
\begin{figure}[!htbp]
    \centering
    \includegraphics[width=0.9\textwidth]{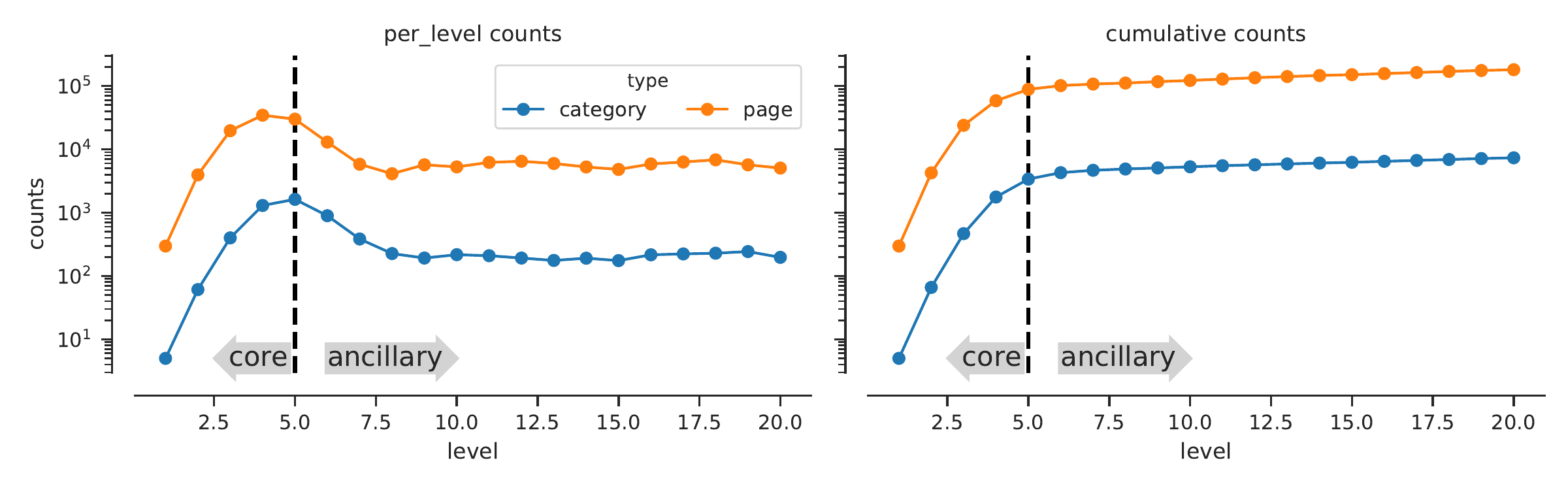}
    \caption{Distribution of subject-heading counts in each level of WikiCSSH}
    \label{fig:wikicssh-dist}
\end{figure}

The final WikiCSSH we built consists of 7K Wikipedia Categories organized as a tree, and 761K affiliated Wikipedia pages. Each category in WikiCSSH has on average 104 affiliated pages. Inter-category parent-child links capture the research field hierarchy, and category-page links capture concepts within a research field. Each category in WikiCSSH is assigned a level based on its lowest identified level in the tree. WikiCSSH contains core CS terms (including categories and pages) in levels 1-7, and ancillary CS terms in levels 8-20 (see figure \ref{fig:wikicssh-dist}). Core terms are highly relevant to CS research topics or concepts, while ancillary terms mainly represent interdisciplinary research topics and concepts. Core terms in WikiCSSH account for 63.5\% of the terms in WikiCSSH. Users of WikiCSSH can decide which part of our vocabulary they want to use depending on their narrow or broad definition of CS and/ or intended use cases. 

\subsection{Evaluation of category extraction based on human annotated data}

\begin{table}[!htbp]
\centering

\caption{Precision (P) and recall (R) in core levels (recall for levels $>5$ cannot be computed as that would require manually annotating all CS-relevant categories.)}
\label{tab:evaluation}
\begin{tabular}{r|rr|rr|rr|rr}
\toprule
  ~ & \multicolumn{2}{c|}{\textbf{\qquad CD}}&
  \multicolumn{2}{c|}{\textbf{\qquad ML}}&
  \multicolumn{2}{c|}{\textbf{\qquad ML+Rule}}&
 \multicolumn{2}{c}{\textbf{CD+ML+Rule}}\\
\midrule
  \textbf{Level} & \textbf{P} & \textbf{R} & \textbf{P} & \textbf{R} & \textbf{P} & \textbf{R} & \textbf{P} & \textbf{R}\\

  \midrule
  1-3 & 1.00 & 1.00 & 1.00 & 1.00 & 1.00 & 1.00 & 1.00 & 1.00\\ 
  4 & 0.36 & 1.00 & 0.64 & 0.88 & 0.85 & 0.90 & 0.85 & 0.88 \\
  5 & 0.17 & 1.00 & 0.66 & 0.90 & 0.87 & 0.79 & 0.87 & 0.79\\
  6 & 0.07 & / & 0.25 & / & 0.47 & / &  0.83 & /\\
  7 & 0.01 & / & 0.23 & / & 0.33 & / & 0.82 & /\\
  \bottomrule
\end{tabular}
\end{table}

This section describes our evaluation of the methods we used for removing false positives. This evaluation also allows us to test if our mixed methods approach can outperform any single method (out of the methods we used) approach to pruning a large-scale dataset with a complex structure such as the WCT. We randomly selected a sample of categories from the ICS before our community detection step, and manually annotated whether the sampled categories were CS-relevant or not. Finally, we leveraged this annotated sample to evaluate precision and recall of different category sets extracted through different methods. It is important to note that we only evaluated categories. Pages inherit the characteristic of relevance to CS from categories they are affiliated with and thus are assumed to share similar results with the evaluation for categories. Table \ref{tab:evaluation} shows the evaluation results. The first three levels are not useful for evaluation as they have been selected manually. From level 4 onward, we find that the embedding based method (ML) achieved a higher precision compared to the community detection (CD) method at the expense of lower recall. Combining ML with rules (Rule) also increases precision at the expense of lower recall, while combining all of CD, ML, and Rule improves the precision significantly in lower levels (more than 0.4 points for levels 6 and 7). This result provides empirical evidence for our argument that mixed methods can outperform a single method approach to prune large-scale data with complex structures.

\section{Comparison and Evaluation} 
To assess the quality of WikiCSSH, this section presents a set of evaluations that are based on comparing WikiCSSH against ACM CCS and CSO with respect to term coverage and phrase extraction performance. Specifically, we first investigate different term coverage patterns of these three vocabularies, and then compare these vocabularies' performance of key-phrase extraction based on abstracts in a scholarly dataset.

\subsection{Comparison against alternative CS vocabularies} 

Table \ref{tab:vocab_summary} shows quantitative statistics for our final vocabulary in comparison to ACM CCS and CSO. WikiCSSH contains $\sim 7.4$ thousand coarse-grained terms (categories) that are associated with $\sim 0.75$ million fine-grained terms (pages) organized into 20 levels. Therefore, WikiCSSH is 375 larger than ACM CCS (around 2,000 terms) and and 33 times larger than CSO (around 14,000 terms). Also, while both ACM CCS and CSO have a hierarchical structure to represent relations between the terms they contain, neither of them distinguishes between coarse-grained (categories) and fine-grained (pages) terms as well as between core and ancillary terms. 

\begin{table}[!htbp]
    \centering
    \caption{Summary of existing subject headings in Computer Science}
    \label{tab:vocab_summary}
    \begin{tabular}{lll}
    \toprule
      \textbf{Vocabulary} & \textbf{Size} & \textbf{Curation}\\ 
      \midrule
      ACM CCS & 2K & Expert Labeling\\ 
      CSO v.3.2 & 14K & Data Mining\\ 
      WikiCSSH & 7.4K categories + 752K pages & Crowdsource + HITL Data Mining\\
      \bottomrule
    \end{tabular}
  \vspace{-0.6cm}
\end{table}

\begin{figure}[!htbp]
    \centering
    \includegraphics[width=0.7\textwidth]{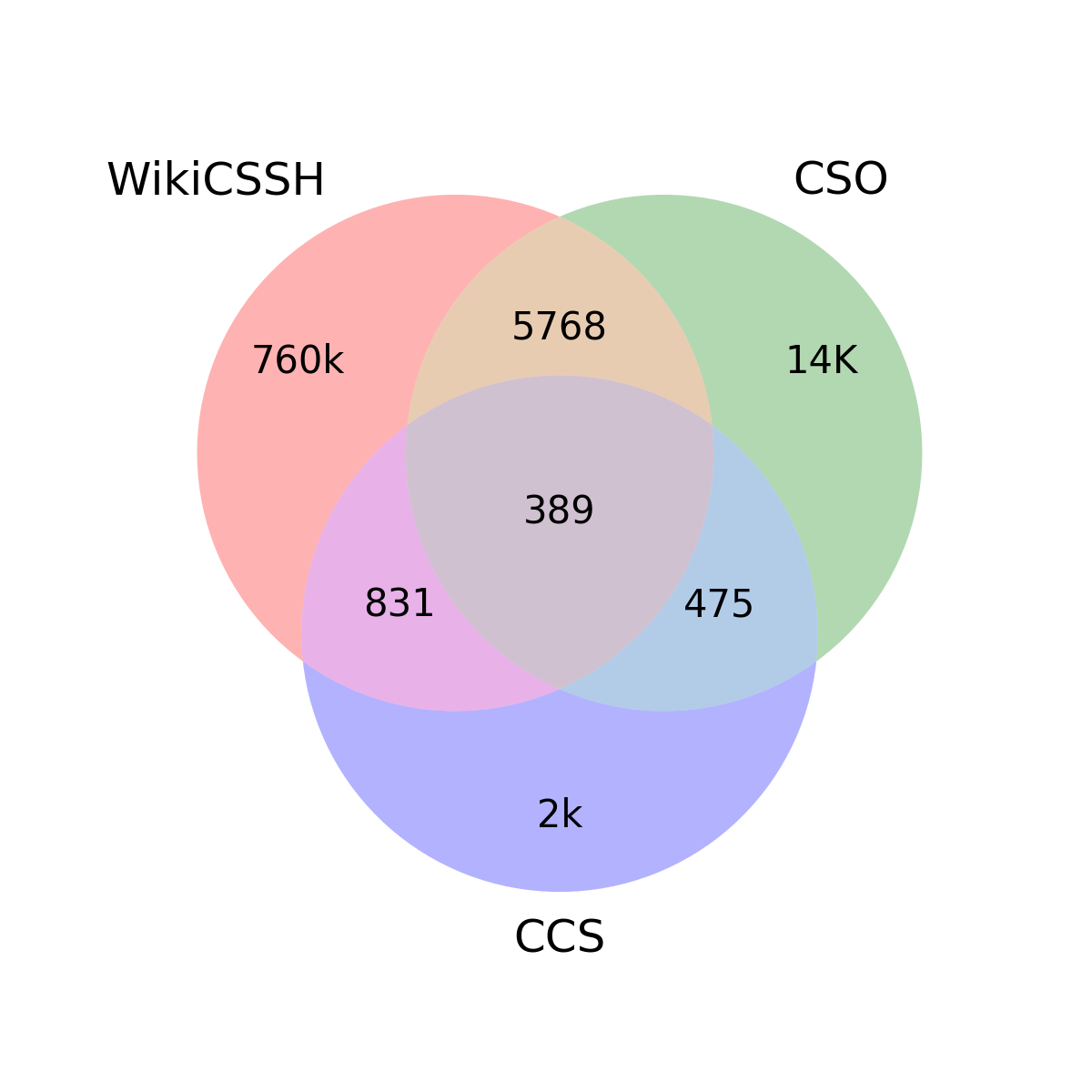}
    \caption{Venn diagram: coverage relations between ACM CCS, CSO, and WikiCSSH}
    \label{fig:venn}
\end{figure}

In addition to vocabulary size, term coverage characteristics are another crucial indicator of vocabulary quality. Figure \ref{fig:venn}, a Venn diagram, shows the relationship between WikiCSSH, CSO, and ACM CCS. Interestingly, none of these vocabularies covers more than half of the lemmatized terms from the other vocabularies. This finding implies that each of these vocabularies is far from a complete list of CS-related terms despite, regardless of their vocabulary size.


\begin{table}[t]
\fontsize{8}{11}\selectfont
\centering
\caption{Example terms in ACM CCS, CSO, and WikiCSSH}
\begin{tabular}{|c|l|}
\hline
\textbf{Venn Relation} & \multicolumn{1}{c|}{\textbf{Examples}} \\ \hline
CCS $\cap$ CSO $\cap$ Wiki & \begin{tabular}[c]{@{}l@{}}parallel algorithm, heuristic evaluation, structured query language, \\stochastic differential equation, mathematical optimization, \\psychology, computability, ray tracing, shortest path, gaussian process \end{tabular}\\ \hline
CCS exclusively &  \begin{tabular}[c]{@{}l@{}}software development process management,
 invariant,\\
 testing with distributed and parallel system,
 column based storage,\\
 formal software verification,
 developmental representation,\\
 protocol testing and verification,
 empirical study in accessibility,\\
 middle box or network appliance,
 early life failure and infant mortality\end{tabular}\\ \hline
CSO exclusively & \begin{tabular}[c]{@{}l@{}}international telecommunication,
 loop closure,
 microwave signal,\\
 speech feature,
 graph matching,
 congenial access control,\\
 biped robot,
 petri net model,
 international,
 semantic data\end{tabular}\\ \hline
Wiki exclusively & \begin{tabular}[c]{@{}l@{}}
 space elevator,
 system of set theory,
  subreddit,
 scientific database,\\
 behavior modelling,
 computer security model,
 euclidean symmetry,\\
 algorithmic information theory,
 dichotomy,
 moment (mathematics)\end{tabular}  \\ \hline
\end{tabular}

\label{tab: keyterms}
\end{table}

A qualitative analysis of the example terms shown in Table \ref{tab: keyterms} further informs us about the different coverage patterns in ACM CCS, CSO, and WikiCSSH. 

\begin{itemize}
    \item \textbf{CCS $\cap$ CSO $\cap$ WikiCSSH:} Unsurprisingly, the vast majority of terms covered by all three vocabularies are short (all terms in the table consist of one to three words), common, and/or referring to high-level concepts or areas. For example, \textit{Gaussian process} is a common term related to modeling, and \textit{parallel algorithm} refers to a broad range of algorithms.
    \\
    \item \textbf{ACM CCS exclusively:} The most prominent characteristic of the terms contained only in ACM CCS is their average length. Specially, the average length of the terms that occur exclusively in ACM CCS, CSO, and WikiCSSH is 3.27, 2.46, and 2.24, respectively. Also, some entries in ACM CCS refer to more than one research area or concept (instead of a concrete single one), such as \textit{middle box or network appliance}. 
    \\
    \item \textbf{CSO exclusively:} The terms exclusively covered by CSO are short and refer to a concrete areas or concepts.
    \\
    \item \textbf{WikiCSSH exclusively:} Many of the terms that occur only in WikiCSSH are also short and referring to concrete areas or concepts. A unique characteristic of terms exclusively in WikiCSSH is their representation of interdisciplinary areas, such as \textit{subreddit} (related to social media studies) and \textit{behavior modelling} (related to computational social science).
    
\end{itemize}

These characteristics also indicate potential advantages and disadvantages with different vocabulary building approaches. 
First, the characteristics of the exclusive terms in ACM CCS represent some of the difficulties with expert-curated vocabularies. One one hand, this type of vocabulary is often in small size. On the other hand, since the intent of ACM CCS is to organize and label research papers into an ontology rather than tagging concrete terms in scholarly work, many of its terms are lengthy and different from the terms used in scholarly writing.

Second, the differences between CSO and WikiCSSH show one of the advantages of utilizing crowd-sourced data to build a vocabulary, i.e., the inclusion of more interdisciplinary terms. Specifically, CSO was constructed by text mining scholarly datasets in core areas of computer science \cite{salatino2018computer}, such as the DBLP dataset \cite{cavacini2015best,osborne2013exploring,petricek2005comparison, reitz2010analysis}. It is unlikely to cover a substantial number of terms related to interdisciplinary areas or concepts.

\subsection{Evaluation of phrase extraction based on human annotated data}

A common application of a domain-specific vocabularies is to tag scholarly papers with domain-specific terms. A vocabulary can be considered effective if it enables the identification of important key-phrases in a dataset of scholarly papers from a given domain. We used this approach, and leveraged the KP20k dataset \cite{gallina2020large,meng2017deep} for our evaluation. The test set in KP20k contains 20,000 CS research abstracts and related, human-annotated key-phrases, such as \emph{machine learning}, \emph{data mining}, and \emph{clustering}.

Before using ACM CCS, CSO, and WikiCSSH to extract phrases from the abstracts in the KP20k corpus, we lemmatized all terms in these vocabularies and all words in the test set of KP20k. Since the vast majority of annotated key-phrases in the KP20k are not unigrams, we removed all (more than 0.1 million) unigram terms from WikiCSSH unless they were covered by CSO or ACM CCS (which suggests that these unigram terms are important). This step substantially reduced the compute time for phrase extraction and increased its precision. After processing the test set and vocabularies, we extracted all exact matches between terms in abstracts in KP20k and any of the three vocabularies. We processed the abstracts one by one (instead of merging them into one text) in order to further analyze which vocabulary can have a more robust and consistent performance across abstracts.

Table \ref{tab: extraction} shows the annotated key-phrases from KP20k and all phrases extracted by the three vocabularies. It is worth noting that the annotated key-phrases do not contain all CS-related phrases in the abstracts; they only represent the most important phrases that summarize the main content of abstracts instead of all CS-related phrases occurring in these abstracts. For instance, in the first example abstract shown in Table \ref{tab: extraction}, the term \textit{local optimum} extracted by WikiCSSH is a CS-related term, but it is not included in the list of annotated key-phrases. However, this test set still allow for evaluation of how much important information these vocabularies can extract.

\begin{landscape}
\begin{table}[t]
\fontsize{6}{9}\selectfont
\centering
\caption{Six examples of phrase extraction results}
\begin{tabular}{|l|l|l|l|}
\hline
\textbf{Annotated Key Phrases} & \textbf{ACM CCS Extraction} & \textbf{CSO Extraction} & \textbf{WikiCSSH Extraction} \\ \hline

\begin{tabular}[c]{@{}l@{}} radio frequency identification  (rfid), \\artificial immune system (ais),\\genetic algorithms (gas),\\fuzzy neural network (fnn) \end{tabular} & \begin{tabular}[c]{@{}l@{}}\textbf{artificial immune system},\\
 evaluation,
 neural network,\\
 supply chain management \end{tabular} & \begin{tabular}[c]{@{}l@{}}frequency, \textbf{fuzzy neural},\\ learning, radio, \\ \textbf{rfid}, warehouse \\neural network\end{tabular} & \begin{tabular}[c]{@{}l@{}} \textbf{artificial immune system},\\
 artificial neural network,\\
 elsevier ltd,
 evaluation,\\
 \textbf{fuzzy neural network},\\
 if then,
  frequency,\\
 immune system,
 local optimum,\\
 memory,
 neural network,\\
   learning,
 radio,
  \textbf{rfid},
 radio frequency,\\
 \textbf{radio frequency identification}\end{tabular}\\ \hline
 
\begin{tabular}[c]{@{}l@{}} tag, concept prediction,\\ large-scale date set, visual feature \end{tabular} &  \begin{tabular}[c]{@{}l@{}}feature selection,\\ annotation,  performance \end{tabular}  &  \begin{tabular}[c]{@{}l@{}}  annotation, probability,\\ \textbf{visual feature}, web image  \end{tabular} & \begin{tabular}[c]{@{}l@{}} average precision,
 cloud distribution,\\
 co occurrence matrix,
 data set,\\
 feature selection,
 for each,\\
 ground truth,
 occurrence matrix,\\
 performance,
 probability,\\
 \textbf{tag cloud} \end{tabular}\\ \hline
 
\begin{tabular}[c]{@{}l@{}} finite element, model updating,\\ structural dynamics, \\ modal strain energy,\\ objective function, vibration test,\\ eigenfrequency, model expansion \end{tabular}& \begin{tabular}[c]{@{}l@{}} No phrase extracted \end{tabular} & \begin{tabular}[c]{@{}l@{}}
 multi objective optimisation,\\
 numerical result,\\
 \textbf{objective function} of,\\
 parameter,
 fe model,\\
 strain energy\end{tabular} & \begin{tabular}[c]{@{}l@{}} algorithm,
case study,\\
 \textbf{finite element}, \\
 multi objective optimisation,\\
 \textbf{objective function},
 parameter,\\
 robust optimisation,\\
 sensitivity study \end{tabular}\\ \hline
 
 \begin{tabular}[c]{@{}l@{}} engineering data modeling,\\ fuzzy express-g data model,\\ fuzzy xml model,\\ transformation  \end{tabular} &  \begin{tabular}[c]{@{}l@{}}  data exchange, \\data modeling, \\ design, engineering, \\markup language \end{tabular} &  \begin{tabular}[c]{@{}l@{}} conceptual design,
 cycle,\\
 data exchange,
 engineering, \\
 graphical representation,\\
 modeling language,\\
 product design,
 xml data,\\
 xml extensible markup language  \end{tabular} & \begin{tabular}[c]{@{}l@{}} conceptual design,
 data exchange,\\
 data model,
 data modeling,\\
 de facto standard,
 \\
 g data,
 markup language,
 express g,\\
 extensible markup language,\\
 graphical representation,\\
 information modeling,\\
 modeling language,\\
 product design,
 design,
 the web,\\
 web based,
 xml data  \end{tabular} \\ \hline
 
\begin{tabular}[c]{@{}l@{}}low noise amplifier,\\ noise figure, cmos, wireless,\\ hearing aid \end{tabular} &  \begin{tabular}[c]{@{}l@{}} printed circuit board \end{tabular}  & \begin{tabular}[c]{@{}l@{}} amplifier, noise \end{tabular} & \begin{tabular}[c]{@{}l@{}} amplifier,
circuit board, \\
\textbf{low noise amplifier},
noise,
noise figure,\\
printed circuit,
printed circuit board \end{tabular}  \\ \hline

\begin{tabular}[c]{@{}l@{}}latent variable model, \\ structural equation model, \\ r, maximum likelihood, \\ serotonin, seasonality, sert \end{tabular} & \begin{tabular}[c]{@{}l@{}} incomplete data, \\
 \textbf{latent variable model},\\
 \textbf{maximum likelihood estimation}, \\
 measurement \end{tabular} & \begin{tabular}[c]{@{}l@{}} extensive simulation,
 inference, \\
 \textbf{latent variable model}, \\
 \textbf{maximum likelihood estimation}, \\
 parameter,
 software \end{tabular} & \begin{tabular}[c]{@{}l@{}} estimation,
 human brain,
 inference,\\
 instrumental variable,
 latent variable, \\
 \textbf{latent variable model},
  measurement, \\
 \textbf{maximum likelihood estimation},\\
 textbf{maximum likelihood},
 non linear, \\
 parameter,
 robust standard error, \\
 \textbf{serotonin} transporter,
 simulation,
 software, \\
 standard error,
 \textbf{structural equation}, \\
 \textbf{structural equation model},
 the human brain \end{tabular} \\ \hline

\end{tabular}
\label{tab: extraction}
\end{table}
\end{landscape}

Our performance evaluation uses a widely used exact match evaluation method \cite{papagiannopoulou2020review, gallina2020large}, where only the extract matches with the gold standard are considered as true positives. Specifically, 
$$Precision = \frac{\sum{Number \ of \ Matched \ in \ Abstract_i}}{\sum{Number \ of \ Extracted \ in \ Abstract_i}},$$
$$Recall = \frac{\sum{Number \ of \ Matched \ in \ Abstract_i}}{\sum{Number \ of \ Annotated \ in \ Abstract_i}}.$$
In this equation, \textbf{Number of Annotated} is the number the annotated key-phrases per abstract, \textbf{Number of Extracted} is the number that a given vocabulary extracted from an abstract, and \textbf{Number of Matched} is the number of extracted phrases that exactly match an annotated key-phrase.

The precision and recall computed based on the exact match evaluation will underestimate the performances of all three vocabularies because these vocabularies can extract name variants of the annotated key-phrases, such as \textit{maximum likelihood} versus \textit{maximum likelihood estimation} as shown in the sixth example in Table \ref{tab: extraction}, but these variations are not considered as correct as per our method. Since more than 50\% of terms in WikiCSSH were from redirect pages (name variants of pages) in Wikipedia, the performance of WikiCSSH may face a high degree of underestimation. However, it is still reasonable to use this stricter evaluation standard to assess WikiCSSH's performance: If WikiCSSH can outperform CSO and ACM CCS under a stricter standard, we can be more confident about claim about its performance.

Table \ref{tab:evaluation} shows the precision, recall, and F-score for ACM CCS, CSO, and WikiCSSH against the gold standard. WikiCSSH has the largest F-score mainly because of its highest recall and acceptable precision: Compared to CSO, WikiCSSH achieves slightly lower precision (-0.008) and higher recall (+0.061). One reason for its slightly lower precision is the characteristic of the terms covered by WikiCSSH. For instance, one phrase extracted by WikiCSSH is \textit{if then}. WikiCSSH contains this term because it is a sub-concept of \textit{conditional (computer programming)}. However, this type of term can cause confusions when it is used to extract phrases from scholarly papers.

\begin{table}[!htbp]
\centering

\caption{Key-phrase extraction performance}
\label{tab:evaluation}
\begin{tabular}{r|c|c|c}
\toprule
  ~ & {\textbf{Precision}}&
  {\textbf{Recall}}&
{\textbf{F1}}\\

  \midrule
  \textbf{ACM CCS} & \textbf{0.100} & 0.037 & 0.054 \\ 
  \textbf{CSO} & 0.077 & 0.068 & 0.072 \\
  \textbf{WikiCSSH} & 0.069 & \textbf{0.129} & \textbf{0.090} \\

  \bottomrule
\end{tabular}
\end{table}

In addition to the corpus-level precision, which indicates overall performance, an investigation of the variation of individual-abstract-level precision can inform us about whether a vocabulary has a consistent and robust performance across abstracts or not. The individual-abstract-level precision can be written as $$Precision_i = \frac{Number \ of \ Matched \ in \ Abstract_i}{Number \ of \ Extracted \ in \ Abstract_i}.$$

\begin{figure}[!htbp]
    \centering
    \includegraphics[width=0.9\textwidth]{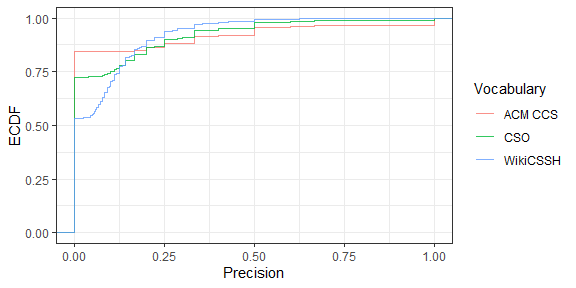}
    \caption{Distribution of individual-abstract-level precision based on empirical cumulative distribution function (Compared to ACM CCS and CSO, WikiCSSH has a relatively smoother line as well as less polarized individual-abstract-level precision that is equal to 0 or larger than 0.5.)}
    \label{fig:precision}
\end{figure}

Figure \ref{fig:precision} shows the distribution of individual-abstract-level precision across abstracts. The standard deviation for these distributions is 0.239 (ACM CCS), 0.170 (CS0), and 0.117 (WikiCSSH), respectively. Interestingly, the precision of WikiCSSH across abstracts is more consistent and robust than that of CSO and ACM CCS, even though its overall precision is lower than theirs. Besides, while CSO and WikiCSSH have a similar overall precision, WikiCSSH performs better than CSO in more abstracts. Specifically, compared to CSO, WikiCSSH performs better in 5233 abstracts and worse in 4535 abstracts. In the rest 10,232 abstracts, they have equal performance because neither of them successfully extracts any exact matches with the gold standard.

Finally, we analyzed the coverage between the gold standard key-phrases against the terms in WikiCSSH, ACM CSS, and CSO, respectively. We did this by searching for all exact matches of lemmatized annotated key-phrases with lemmatized terms from the studied vocabularies. Instead of using only the test set in KP20k, we merged the test, validation, and training sets (around 570,000 abstracts with more than 3 million annotated key-phrases). For our evaluation, we counted both the number of unique matched phrases and the total number of matched phrases. Since the total number of phrase matches is going to be biased towards frequently occurred concepts that are likely to be present in all vocabularies, we also used unique phrase matches to identify coverage. As reported in table \ref{tab:evaluation_kp20k}, WikiCSSH extracted 62,142 (8.7\%) unique phrases and 1,257,731 (41.7\%) total phrases from KP20k, and most of them were from WikiCSSH's core part. The numbers of extracted unique and total phrases for ACM were 1,266 (0.18\%) and 272,105 (9\%), and for CSO 11,256 (1.58\%) and 733,071 (24.3\%). This evaluation suggests that WikiCSSH supports comparatively high coverage of CS key-phrases from scholarly texts. 

\begin{table}[!htbp]
\centering
\caption{Comparison of coverage of various vocabularies on annotated keyphrases in KP20k corpus (percents = phrases matched by vocabulary / annotated phrases in KP20K)}
\label{tab:evaluation_kp20k}

\begin{tabular}{
>{\raggedright\arraybackslash}p{0.28\textwidth}%
>{\raggedleft\arraybackslash}p{0.24\textwidth}%
>{\raggedleft\arraybackslash}p{0.24\textwidth}%
}
\toprule
  \textbf{Vocabulary} & \textbf{Unique Phrases} & \textbf{Total Phrases}\\
  \midrule
  ACM CCS & 1,266 (0.18\%) & 272,105 (9\%) \\ 
  CSO & 11,265 (1.58\%) & 733,071 (24.3\%) \\
  WikiCSSH (core) & 44,974 (6.3\%) & 1,031,156 (34.2\%)\\
  WikiCSSH (ancillary) & 17,168 (2.4\%) & 226,575 (7.5\%)\\
  WikiCSSH (total)& 62,142 (8.7\%) & 1,257,731 (41.7\%)\\
  \bottomrule
\end{tabular}
\end{table}

We further calculated the ratios of total to unique phrases, respectively, for the core and ancillary part of the WikiCSSH, which show WikiCSSH's ability to extract rare phrases from KP20K. We found that the ratio of total to unique phrases for the core part of WikiCSSH is 22.93, while for the ancillary part, it is only 13.2. Put differently, the core part of WikiCSSH is more likely to capture frequently occurring phrases in CS research articles, while the ancillary part is more to capture rare phrases. Similarly, for ACM CCS and CSO, the ratio of total to unique phrases were 214.93 and 65.08, respectively. This result indicates that WikiCSSH is more likely to extract rare phrases from scholarly articles compared to ACM CCS and CSO. CSO contains a lower proportion of rare phrases that occur in KP20K compared to WikiCSSH. A possible reason for this lower coverage may be the inability of automated data mining methods  to capture low probability signals.

\subsection{Application to identifying keyphrases and salient categories in abstracts}

\begin{figure}
    \centering
    \includegraphics[width=\textwidth]{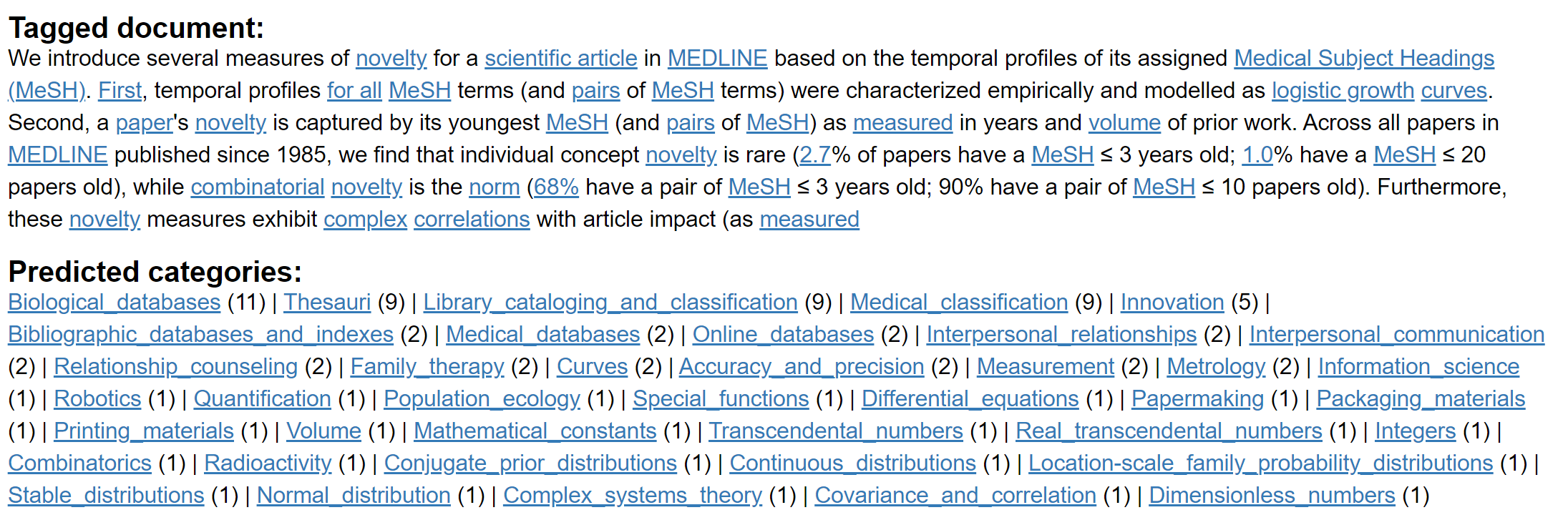}
    \caption{A bio-bibliometrics paper abstract tagged using WikiCSSH as well as category counts from WikiCSSH}
    \label{fig:tagging}
\end{figure}

In figure \ref{fig:tagging} we present an example of a bio-bibliometrics paper \cite{Mishra2016a} abstract tagged using key-phrases identified via Wikidata \footnote{Code at: \url{https://github.com/uiuc-ischool-scanr/WikiCSSH/blob/master/notebooks/Tagging_using_WikiCSSH.ipynb}}. We also show the category counts as identified using WikiCSSH. These category counts can help us know the key topics of a given paper. This demonstrates a good usage of WikiCSSH as a crude measure of identifying the key topical categories of a paper by limiting outselves to the top few categories. 

\section{Conclusion, Discussion and Limitations}

We have presented WikiCSSH, a large-scale subject headings vocabulary for the CS domain. We have developed WikiCSSH using a human-in-the-loop workflow that leverages crowd-sourced Wikipedia data. WikiCSSH outperforms two alternative CS vocabularies, namely ACM CCS and CSO, in number of items, coverage of key-phrases in a benchmark dataset of scholarly papers from CS, and performance of key-phrase extraction from scholarly abstracts. Users of WikiCSSH can decide which part of WikiCSSH they want to use depending on their needs. For example, users may want to i) use the 7,355 hierarchically structured categories for indexing (research areas and topics in) documents, or ii) use the 0.75 million concrete, fine-grained terms (from pages) within categories for more detailed concept analysis, or iii) select the core and/or ancillary part of WikiCSSH according to their broad or narrow definition of computer science as needed for their work. 

Our work contributes to methodological insights about how to leverage crowd-sourced data when the main challenge is to prune false positives to increase precision for target applications. Building a sizeable and domain-specific vocabulary like WikiCSSH would be extremely expensive and/ or time consuming if one had to rely on manual work by domain experts. However, existing crowd-sourced data with a permissible license opens up an opportunity to build a large-scale, structured vocabulary at low cost in terms of both time and human resources. That being said, our approach is more costly and time-consuming than a fully automated data mining-based approach due to the substantial human interventions that we made part of our process. However, we showed that our approach can capture relevant yet rare phrases that might be ignored by fully automated data mining solutions. Our work also illustrates the challenges resulting from using the given structure of Wikipedia data for our specific task and assesses possible solutions to overcome these challenges through the methods described in this paper. Our workflow can be extended to construct subject headings for other domains by modifying the rules and training approaches. Code for replicating the construction and refinement of WikiCSSH along with the latest version of WikiCSSH can be found at: \href{https://github.com/uiuc-ischool-scanr/WikiCSSH}{https://github.com/uiuc-ischool-scanr/WikiCSSH} \cite{WikiCSSH}.

Finally, our comparative analysis shows that none of the three studied vocabularies, which were built with different approaches (i.e., expert curation for ACM CCS, data mining of scholarly work for CSO, and crowd-sourcing for WikiCSSH) is complete or can comprehensively outperform any other vocabulary. More research is needed to build a more complete vocabulary. Therefore, we acknowledge the limitations of WikiCSSH as well as our vocabulary building approach. In our future work, we plan to update WikiCSSH based on new data and with new methods, further remove noisy terms, and test its performance for other tasks such as indexing and categorization.

\bibliographystyle{splncs04}
\bibliography{references}

\begin{thebibliography}{10}
\providecommand{\url}[1]{\texttt{#1}}
\providecommand{\urlprefix}{URL }
\providecommand{\doi}[1]{https://doi.org/#1}

\bibitem{blondel2008fast}
Blondel, V.D., Guillaume, J.L., Lambiotte, R., Lefebvre, E.: Fast unfolding of
  communities in large networks. Journal of Statistical Mechanics: Theory and
  Experiment  \textbf{2008}(10),  P10008 (oct 2008).
  \doi{10.1088/1742-5468/2008/10/p10008}

\bibitem{cavacini2015best}
Cavacini, A.: What is the best database for computer science journal articles?
  Scientometrics  \textbf{102}(3),  2059--2071 (2015)

\bibitem{gallina2020large}
Gallina, Y., Boudin, F., Daille, B.: Large-scale evaluation of keyphrase
  extraction models. In: Proceedings of the ACM/IEEE Joint Conference on
  Digital Libraries in 2020. pp. 271--278 (2020)

\bibitem{girvan2002community}
Girvan, M., Newman, M.E.: Community structure in social and biological
  networks. Proceedings of the national academy of sciences  \textbf{99}(12),
  7821--7826 (2002)

\bibitem{Grover2016}
Grover, A., Leskovec, J.: Node2vec: Scalable feature learning for networks. In:
  Proceedings of the 22nd ACM SIGKDD International Conference on Knowledge
  Discovery and Data Mining. p. 855–864. KDD ’16, Association for Computing
  Machinery, New York, NY, USA (2016). \doi{10.1145/2939672.2939754}

\bibitem{WikiCSSH}
Han, K., Yang, P., Mishra, S., Diesner, J.: Wikicssh - computer science subject
  headings from wikipedia (2020). \doi{10.13012/B2IDB-0424970\_V1}

\bibitem{WikiCSSH2020Han}
Han, K., Yang, P., Mishra, S., Diesner, J.: Wikicssh: Extracting computer
  science subject headings from wikipedia. In: Bellatreche, L., Bielikov{\'a},
  M., Boussa{\"i}d, O., Catania, B., Darmont, J., Demidova, E., Duchateau, F.,
  Hall, M., Mer{\v{c}}un, T., Novikov, B., Papatheodorou, C., Risse, T.,
  Romero, O., Sautot, L., Talens, G., Wrembel, R., {\v{Z}}umer, M. (eds.)
  ADBIS, TPDL and EDA 2020 Common Workshops and Doctoral Consortium. pp.
  207--218. Springer International Publishing, Cham (2020)

\bibitem{hoffart2013yago2}
Hoffart, J., Suchanek, F.M., Berberich, K., Weikum, G.: Yago2: A spatially and
  temporally enhanced knowledge base from wikipedia. Artificial Intelligence
  \textbf{194},  28 -- 61 (2013).
  \doi{https://doi.org/10.1016/j.artint.2012.06.001}

\bibitem{lehmann2015dbpedia}
Lehmann, J., Isele, R., Jakob, M., Jentzsch, A., Kontokostas, D., Mendes, P.N.,
  Hellmann, S., Morsey, M., Van~Kleef, P., Auer, S., et~al.: Dbpedia--a
  large-scale, multilingual knowledge base extracted from wikipedia. Semantic
  Web  \textbf{6}(2),  167--195 (2015)

\bibitem{levine2010rankings}
Levine, T.R.: Rankings and trends in citation patterns of communication
  journals. Communication Education  \textbf{59}(1),  41--51 (2010)

\bibitem{medelyan2008topic}
Medelyan, O., Witten, I.H., Milne, D.: Topic indexing with wikipedia. In:
  Proceedings of the AAAI WikiAI workshop. vol.~1, pp. 19--24 (2008)

\bibitem{meng2017deep}
Meng, R., Zhao, S., Han, S., He, D., Brusilovsky, P., Chi, Y.: Deep keyphrase
  generation. arXiv preprint arXiv:1704.06879  (2017)

\bibitem{Mishra2018Expertise}
Mishra, S., Fegley, B.D., Diesner, J., Torvik, V.I.: {Expertise as an aspect of
  author contributions}. In: Workshop on Informetric and Scientometric Research
  (SIG/MET). Vancouver (2018)

\bibitem{Mishra2018SelfCite}
Mishra, S., Fegley, B.D., Diesner, J., Torvik, V.I.: {Self-citation is the
  hallmark of productive authors, of any gender}. PLOS ONE  \textbf{13}(9),
  e0195773 (sep 2018). \doi{10.1371/journal.pone.0195773}

\bibitem{Mishra2016a}
Mishra, S., Torvik, V.I.: {Quantifying Conceptual Novelty in the Biomedical
  Literature.} D-Lib magazine : the magazine of the Digital Library Forum
  \textbf{22}(9-10) (sep 2016). \doi{10.1045/september2016-mishra}

\bibitem{Nickel2017}
Nickel, M., Kiela, D.: {Poincar{\'{e}} Embeddings for Learning Hierarchical
  Representations}. In: Advances in Neural Information Processing Systems 30.
  pp. 6338--6347. Curran Associates, Inc. (2017)

\bibitem{Nielsen2017Scholia}
Nielsen, F.{\AA}., Mietchen, D., Willighagen, E.: Scholia, scientometrics and
  wikidata. In: The Semantic Web: ESWC 2017 Satellite Events. pp. 237--259.
  Springer International Publishing, Cham (2017)

\bibitem{osborne2015klink}
Osborne, F., Motta, E.: Klink-2: integrating multiple web sources to generate
  semantic topic networks. In: International Semantic Web Conference. pp.
  408--424. Springer (2015)

\bibitem{osborne2013exploring}
Osborne, F., Motta, E., Mulholland, P.: Exploring scholarly data with rexplore.
  In: International semantic web conference. pp. 460--477. Springer (2013)

\bibitem{Packalen2019}
Packalen, M., Bhattacharya, J.: Age and the trying out of new ideas. Journal of
  Human Capital  \textbf{13}(2),  341--373 (2019). \doi{10.1086/703160}

\bibitem{papagiannopoulou2020review}
Papagiannopoulou, E., Tsoumakas, G.: A review of keyphrase extraction. Wiley
  Interdisciplinary Reviews: Data Mining and Knowledge Discovery
  \textbf{10}(2),  e1339 (2020)

\bibitem{Peters2018}
Peters, M., Neumann, M., Iyyer, M., Gardner, M., Clark, C., Lee, K.,
  Zettlemoyer, L.: {Deep Contextualized Word Representations}. In: Proceedings
  of the 2018 Conference of the NAACL-HLT. pp. 2227--2237. Association for
  Computational Linguistics, Stroudsburg, PA, USA (jun 2018).
  \doi{10.18653/v1/N18-1202}

\bibitem{petricek2005comparison}
Petricek, V., Cox, I.J., Han, H., Councill, I.G., Giles, C.L.: A comparison of
  on-line computer science citation databases. In: International Conference on
  Theory and Practice of Digital Libraries. pp. 438--449. Springer (2005)

\bibitem{reitz2010analysis}
Reitz, F., Hoffmann, O.: An analysis of the evolving coverage of computer
  science sub-fields in the dblp digital library. In: International Conference
  on Theory and Practice of Digital Libraries. pp. 216--227. Springer (2010)

\bibitem{salatino2018computer}
Salatino, A.A., Thanapalasingam, T., Mannocci, A., Osborne, F., Motta, E.: The
  computer science ontology: a large-scale taxonomy of research areas. In:
  International Semantic Web Conference. pp. 187--205 (2018)

\bibitem{shang2018automated}
Shang, J., Liu, J., Jiang, M., Ren, X., Voss, C.R., Han, J.: Automated phrase
  mining from massive text corpora. IEEE Transactions on Knowledge and Data
  Engineering  \textbf{30}(10),  1825--1837 (2018)

\bibitem{wang2010timely}
Wang, Y., Zhu, M., Qu, L., Spaniol, M., Weikum, G.: Timely yago: harvesting,
  querying, and visualizing temporal knowledge from wikipedia. In: Proceedings
  of the 13th International Conference on Extending Database Technology. pp.
  697--700 (2010)

\end{thebibliography}

\end{document}